\documentclass[11pt,a4paper]{article}
\usepackage[colorlinks=true,linkcolor=blue]{hyperref}
\usepackage{cmap}
\usepackage[utf8x]{inputenc}
\usepackage[T1]{fontenc}
\usepackage[english]{babel}
\usepackage{amssymb,amsmath}
\usepackage{amsthm}
\usepackage{bbm}
\usepackage{microtype}
\usepackage{lmodern}

{
\rmfamily
\DeclareFontShape{T1}{lmr}{m}{sc}{<->ssub*cmr/m/sc}{}
\DeclareFontShape{T1}{lmr}{b}{sc}{<->ssub*cmr/b/sc}{}
\DeclareFontShape{T1}{lmr}{bx}{sc}{<->ssub*cmr/bx/sc}{}
}

\newcommand{\thmheadercommand}[1]{\textbf{\scshape{}#1.\\*}}

\newtheoremstyle{yannthm}{\topsep}{\topsep}{\slshape}{}{\scshape\bfseries}{.}{.5em}{%
\thmname{#1}\thmnumber{ #2}\thmnote{#3}%
}
\newtheoremstyle{yannthm2}{\topsep}{\topsep}{}{}{\scshape\bfseries}{.}{.5em}{%
\thmname{#1}\thmnumber{ #2}\thmnote{#3}%
}

\def\d{\operatorname{d}\!{}}

\def\N{{\mathbb{N}}}

\newcommand{\deq}{\mathrel{\mathop:}=}

\newcommand{\from}{\colon} 

\def\eps{\varepsilon}
\renewcommand{\epsilon}{\varepsilon}
\renewcommand{\phi}{\varphi}

\let\oldPr\Pr
\renewcommand{\Pr}{\oldPr\nolimits}
\newcommand{\E}{\mathbb{E}}

\DeclareMathOperator{\Tr}{Tr}

\newcommand{\norm}[1]{\left\|#1\right\|}

\DeclareMathOperator*{\argmax}{arg\,max}

\newenvironment{dem}[1][]{\begin{proof}[\thmheadercommand{Proof#1}]~\newline\ignorespaces}{\end{proof}}

{
\theoremstyle{yannthm}
\newtheorem{defi}{Definition}
\newtheorem*{defi*}{Definition}
\newtheorem{prop}[defi]{Proposition}
\newtheorem*{prop*}{Proposition}
\newtheorem{thm}[defi]{Theorem}
\newtheorem*{thm*}{Theorem}

\newtheorem*{lem*}{Lemma}
\newtheorem{cor}[defi]{Corollary}
\newtheorem*{cor*}{Corollary}

\newtheorem*{ex*}{Example}

\newtheorem*{subenonce}{}

\theoremstyle{yannthm2}

\newtheorem*{exo*}{Exercise}

\newtheorem*{rem*}{Remark}

\newtheorem*{subenonce2}{}

}

\newcommand{\transp}[1]{#1^{\!\top}\!}

\usepackage{url}

\newcommand{\plugin}{p^\mathrm{ML}}
\newcommand{\instNML}{p^\mathrm{SNML}}
\newcommand{\wNML}[1]{p^{#1\text{-}\mathrm{SNML}}}

\newcommand{\thetaML}{\theta^\mathrm{ML}}
\newcommand{\thetaMLp}[1]{\theta^{\mathrm{ML}+#1}}
\newcommand{\fnorm}[1]{\norm{#1}_F}
\newcommand{\fscal}[2]{\left\langle#1\middle.,#2\right\rangle_F}
\newcommand{\thetalap}{{\theta^{\mathrm{Lap}}}}
\newcommand{\I}{\mathcal{I}}

\title{Laplace's rule of succession in information geometry}
\author{Yann Ollivier}
\date{}

\begin{document}
\maketitle

When observing data $x_1,\ldots,x_t$ modelled by a probabilistic
distribution $p_\theta(x)$, the maximum likelihood (ML) estimator
$\thetaML=\argmax_\theta \sum_{i=1}^t \ln p_\theta(x_i)$ cannot, in
general, safely be used to predict $x_{t+1}$. For instance, for a
Bernoulli process, if only ``tails'' have been observed so far, the
probability of ``heads'' is estimated to $0$. Laplace's famous ``add-one'' rule of
succession (e.g., \cite{GrunwaldMDL}) regularizes $\theta$ by adding $1$ to the count of
``heads'' and of ``tails'' in the observed sequence.

Bayesian estimators suffer less from this problem, as every value of
$\theta$ contributes, to some extent, to the Bayesian prediction of
$x_{t+1}$ knowing $x_{1:t}$. However, their use can be limited by the need to
integrate over parameter space or to use Monte Carlo samples from the
posterior distribution.

For Bernoulli distributions, Laplace's rule is equivalent to using a
uniform prior on the Bernoulli parameter. The non-informative Jeffreys
prior on the Bernoulli parameter corresponds to Krichevsky and Trofimov's
``add-one-half'' rule \cite{KT1981}. Thus, in this case, some Bayesian predictors have
a simple implementation.

We claim (Theorem~\ref{thm:main}) that for exponential
families\footnote{For simplicity we only state the results
with i.i.d.\ models.
However the ideas extend to non-i.i.d.\ sequences with
$p_\theta(x_{t+1}|x_{1:t})$ in an exponential family, e.g., Markov
models.}, Bayesian predictors can be approximated by
mixing the ML estimator with the \emph{sequential normalized maximum likelihood}
(SNML) estimator from universal coding theory
\cite{Roos_NML,RoosRissanen_SNML}, which is a fully
canonical version of Laplace's rule.  The weights of
this mixture depend on the density of the desired Bayesian prior with
respect to the non-informative Jeffreys prior, and are equal to $1/2$ for
the Jeffreys prior, thus extending Krichevsky and Trofimov's result. The
resulting mixture also approximates the ``flattened'' ML estimator from
\cite{Grunwald_flattened}.

Thus, it is possible to approximate Bayesian predictors without the cost
of
integrating over $\theta$ or sampling from the posterior. The statements
below emphasize the special role of the Jeffreys prior and the Fisher
information metric. Moreover, the analysis reveals that the direction of
the shift from the ML predictor to Bayesian predictors is systematic and
given by an intrinsic, information-geometric vector field on statistical
manifolds. This could contribute to regularization procedures in
statistical learning.

\paragraph{Notation and statement.}
Let $p_\theta(x)$ be a family of distributions on a variable $x$, smoothly parametrized by $\theta$. Let
$x_1,\ldots,x_t$ be a sequence of observations to be
predicted online using $p_\theta$.
The maximum likelihood predictor is
\begin{equation}
\plugin(x_{t+1}=y|x_{1:t})\deq p_{\thetaML_t}(y)
,\qquad
\thetaML_t\deq \argmax_\theta \sum_{i=1}^t \ln p_\theta(x_i)
\end{equation}
Bayesian predictors (e.g., Laplace's rule) usually differ from $\plugin$ at order
$1/t$.

The \emph{sequential normalized maximum likelihood} predictor
\cite{Roos_NML,RoosRissanen_SNML}
uses,
for each possible value $y$ of $x_{t+1}$, the parameter
$\thetaMLp{y}$ that would yield the best probability if $y$ had already
been observed. Since this increases the probability of every $y$, it is
necessary to renormalize. Define
\begin{equation}
\thetaMLp{y}_t\deq \argmax_\theta \left\{\ln p_\theta(y)+\sum_{i=1}^t \ln
p_\theta(x_i)\right\}
\end{equation}
as the ML estimator when adding $y$ at position $t+1$. For each $y$ let
\begin{equation}
\instNML(x_{t+1}=y|x_{1:t})\deq \frac{1}{Z} p_{\thetaMLp{y}_t}(y)
\end{equation}
be the SNML predictor for time $t+1$, where $Z$ is a normalizing
constant.\footnote{This variant of SNML is SNML-1 in \cite{Roos_NML} and
CNML-3 in \cite{GrunwaldMDL}.}

For Bernoulli distributions, $\instNML$ coincides with Laplace's
``add-one'' rule.\footnote{Note that we describe it in a different way.
The usual presentation of Laplace's rule is to define $\thetalap\deq
\argmax_\theta \{ \ln p_\theta(\text{``heads''})+\ln
p_\theta(\text{``tails''})+\sum \ln p_\theta(x_i)\}$ and then use
$\thetalap$ to predict $x_{t+1}$. Here we follow the SNML viewpoint and use a different $\thetaMLp{y}$ for
each possible value $y$ of $x_{t+1}$.}
For other distributions the two may differ: for
instance, defining Laplace's rule
for continuous-valued $x$ requires choosing a prior
distribution on $x$, whereas the SNML distribution is completely canonical.

We claim that for exponential families, $\frac{1}{2} (\plugin+\instNML)$
is close to the Bayesian predictor using the Jeffreys prior. This
generalizes the ``add-one-half'' rule.

This extends to any Bayesian prior $\pi$ by using a \emph{weighted}
SNML predictor
\begin{equation}
\wNML{w}(y)\deq \frac1Z w(\thetaMLp{y})\, p_{\thetaMLp{y}}(y)
\end{equation}
The weight $w(\theta)$ to be used for a given prior $\pi$ will depend on the ratio
between $\pi$ and the Jeffreys prior. Recall that the latter is
$\pi^{\mathrm{Jeffreys}}(\d\theta)\deq \sqrt{\det \I(\theta)}\d\theta$
where $\I$ is the \emph{Fisher information matrix} of the family
$(p_\theta)$,
\begin{equation}
\I(\theta)\deq -\E_{x\sim p_\theta} \partial^2_\theta \ln p_\theta(x)
\end{equation}
where $\partial^2_\theta$ stands for the Hessian matrix of a function
of $\theta$.

\begin{thm}
\label{thm:main}
Let $p_\theta$ be an exponential family of probability distributions, and
let
$\pi$ be a Bayesian prior on $\theta$.
Then, under suitable regularity assumptions, the Bayesian predictor with
prior $\pi$ knowing $x_{1:t}$
is equal to
\begin{equation}
\frac12 \plugin(\cdot|x_{1:t})+\frac12 \wNML{\beta^2}(\cdot|x_{1:t})
\end{equation}
up to $O(1/t^2)$, where
$\beta(\theta)$ is the density of $\pi$ with respect to the
Jeffreys prior, i.e., $\pi(\d\theta)=\beta(\theta)\sqrt{\det
\I(\theta)}\d\theta$ with $\I$ the Fisher matrix.

More precisely, both under the prior $\pi$ and under
$\frac12(\plugin+\wNML{\beta^2})$, the probability that $x_{t+1}=y$ given
$x_{1:t}$ is asymptotically
\begin{equation}
\label{eq:explicit}
p_{\thetaML_t}(y)\left(1+
\frac1{2t} \fnorm{\partial_\theta \ln p_\theta(y)}^2
+\frac1t \fscal{\partial_\theta \ln \beta}{\partial_\theta \ln
p_\theta(y)}
-\frac{\dim \Theta}{2t}
+O(1/t^2)
\right)
\end{equation}
provided $p_{\thetaML_t}(y)>0$,
where $\fscal{\partial_\theta f}{\partial_\theta g}\deq
\transp{(\partial_\theta
f)}\I^{-1}(\theta)\partial_\theta g$ is the Fisher scalar product and
$\fnorm{\partial_\theta f}^2 = \fscal{\partial_\theta f}{\partial_\theta
f}$
is the Fisher metric norm
of $\partial_\theta f$.
\end{thm}

%

For the Jeffreys prior (constant $\beta$), this also coincides up to
$O(1/t^2)$ with the ``flattened'' or ``squashed'' ML predictor from
\cite{Grunwald_flattened,Grunwald_wrongmodel} with $n_0=0$.  In particular, the latter is
$O(1/t^2)$ close to the Jeffreys prior, and the optimal
regret guarantees
in \cite{Grunwald_flattened} apply to
$\eqref{eq:explicit}$.
Note that a multiplicative $1+O(1/t^2)$
difference between predictors results in an $O(1)$ difference on
cumulated regrets.

\paragraph*{Regularity assumptions.} In most of the article we assume that $p_\theta(x_{t+1}|x_{1:t})$ is a
non-degenerate exponential family of probability distributions. The key
property we need from exponential families is the existence of a
parametrization $\theta$ in which $\partial^2_\theta \ln
p_\theta(x)=-\I(\theta)$ for all $x$ and $\theta$.
For simplicity we assume that the space for $x$ is compact, so
that to prove $O(1/t^2)$ convergence of distributions over $x$ it is
enough to prove $O(1/t^2)$ convergence for each value
of $x$.
We assume that the sequence of
observations $(x_t)_{t\in\N}$ is an \emph{ineccsi sequence}
\cite{GrunwaldMDL}, namely, that for $t$ large
enough, the maximum likelihood estimate stays in a compact
subset of the parameter space.
The Bayesian priors are assumed to be smooth with positive
densities.
In some parts of the article we do not need $p_\theta$ to be an
exponential family, but we still assume that the model $p_\theta$ is smooth,
that there is a
well-defined maximum $\thetaML_t$ for any $x_{1:t}$ and no other
log-likelihood local maxima.

\paragraph{Computing the SNML predictor.} We prove Theorem~\ref{thm:main}
by proving that both predictors are given by~\eqref{eq:explicit}.
Further proofs are gathered at the end of the text.

We
first work on $\instNML$.  Here we do not assume that $p_\theta$ is an
exponential family.  
Let $J_t$ be 
the \emph{observed information matrix}, assumed to be positive-definite,
\begin{equation}
J_t(\theta)\deq -\frac1t \sum_{i=1}^t \partial^2_\theta \ln p_\theta(x_i)
\end{equation}

\begin{prop}
\label{prop:onlineML}
Under suitable regularity assumptions, the maximum likelihood update from
$t$ to $t+1$ satisfies
\begin{equation}
\thetaML_{t+1}= \thetaML_t+\frac1t J_t(\thetaML_t)^{-1}\,\partial_\theta \ln
p_\theta(x_{t+1})+O(1/t^2)
\end{equation}
\end{prop}

For exponential families, this update is the
natural gradient of $\ln p(x_{t+1})$ with learning rate $1/t$
\cite{Amari1998},
because $J_t(\thetaML_t)=\I(\thetaML_t)$, the exact Fisher information
matrix. (For exponential families \emph{in the natural
parametrization}, $J_t(\theta)=\I(\theta)$ for
all $\theta$. But since the Hessian of a function $f$ on a manifold is a
well-defined tensor at a critical point of $f$, it follows that at
$\thetaML_t$ one has $J_t(\thetaML_t)=\I(\thetaML_t)$ for \emph{any}
parametrization of an exponential family.)

\begin{prop}
\label{prop:NML}
Under suitable regularity assumptions,
\begin{equation}
\instNML(y|x_{1:t})=\frac1Z\,
p_{\thetaML_t}(y)
\left(
1+\frac1t \transp{(\partial_\theta \ln p_\theta(y))}J_t^{-1}\,\partial_\theta \ln
p_\theta(y)
+O(1/t^2)
\right)
\end{equation}
provided $p_{\thetaML_t}(y)>0$, where $J_t$ is as above and
the derivatives are taken at $\thetaML_t$.
\end{prop}

Importantly, the normalization constant $Z$ can be computed without
having to sum over $y$ explicitly. Indeed
(cf.~\cite{Grunwald_flattened}), by definition of $\I(\theta)$,
\begin{equation}
\E_{y\sim p_\theta}
\transp{(\partial_\theta \ln p_\theta(y))}J_t^{-1}\partial_\theta\ln
p_\theta(y)=\Tr(J_t^{-1}\I(\theta))
\end{equation}
so that
$Z=1+\frac1t \Tr (J_t^{-1}\I(\thetaML_t))+O(1/t^2)$. For exponential
families,
$J_t=\I$ at $\thetaML_t$ so that $Z=1+\frac{\dim \Theta}{t}+O(1/t^2)$ and
\begin{equation}
p_{\thetaML_t}(y)
\left(
1+\frac1t \transp{(\partial_\theta \ln
p_\theta(y))}\,\I^{-1}\,\partial_\theta \ln
p_\theta(y)-\frac{\dim \Theta}{t}\right)
\end{equation}
is an $O(1/t^2)$ approximation of $\instNML(y|x_{1:t})$.

For the weighted SNML distribution $\wNML{w}$, a similar argument yields
\begin{equation}
\label{eq:wNML}
\wNML{w}(y|x_{1:t})=\frac1Z p_{\thetaML_t}(y)\left(
1+\frac1t
\transp{(\partial_\theta \ln
p_\theta(y))}J_t^{-1}\left(\partial_\theta \ln
p_\theta(y)
+
\partial_\theta \ln
w(\theta)\right)
+O(1/t^2)
\right)
\end{equation}
with $Z=1+\frac1t \Tr (J_t^{-1}\I(\thetaML_t))+O(1/t^2)$ as above.
(The
$\partial_\theta\ln w$ term does not contribute to $Z$ because $\sum_y p_\theta(y)\partial_\theta \ln p_\theta(y)=0$.)
%

Computing $\frac12\plugin+\frac12\wNML{w}$ with $w(\theta)=\beta(\theta)^2$ in
\eqref{eq:wNML}, and using that $J_t(\thetaML)=\I$ for exponential
families, proves one half of Theorem~\ref{thm:main}.

\paragraph{Computing the Bayesian posterior.} Next, let us establish the
asymptotic behavior of the Bayesian posterior. This relies on results
from \cite{TierneyKadane1986}.
The following proposition
may have independent interest.

\begin{prop}
\label{prop:bayes}
Consider a Bayesian prior $\pi(\d\theta)=\alpha(\theta)\d\theta$. Then
the posterior mean of a smooth function $f(\theta)$ given
data $x_{1:t}$ and prior $\pi$ is
asymptotically
\begin{equation}
\label{eq:posterior}
f(\thetaML_t)+
\frac1t
\transp{(\partial_\theta f)}J_t^{-1}\partial_\theta\left(\ln
\frac{\alpha}{\sqrt{\det (-\partial^2_\theta L)}}\right)
+ \frac1{2t} \Tr (J_t^{-1} \partial^2_\theta f)+O(1/t^2)
\end{equation}
where $L(\theta)\deq \frac1t \ln p_\theta(x_{1:t})$ is the average log-likelihood
function, $\partial^2_\theta$ is the Hessian matrix w.r.t.\ $\theta$, and $J_t\deq -\partial^2_\theta L(\thetaML_t)$ is the
observed information matrix.
\end{prop}

When $p_\theta$ is an exponential family in
the natural parametrization, for any $x_{1:t}$, $-\partial^2_\theta L$ is equal to
the Fisher matrix $\I$, so that the denominator in the
log is the Jeffreys prior $\sqrt{\det \I}$. In particular, for
exponential families in natural coordinates, the first term
vanishes if the prior $\pi$ is
the Jeffreys prior.

\begin{cor}
\label{cor:exp}
Let $p_\theta$ be an exponential family.
Consider a Bayesian prior
$\beta(\theta)\sqrt{\det \I(\theta)}\d\theta$ having density $\beta$ with respect
to the Jeffreys prior. Then the posterior probability that $x_{t+1}=y$ knowing
$x_{1:t}$ is asymptotically given by \eqref{eq:explicit} as in
Theorem~\ref{thm:main}.
\end{cor}

This proves the second half of Theorem~\ref{thm:main}.

\paragraph{Intrinsic viewpoint.} When rewritten in intrinsic Riemannian
terms, Proposition~\ref{prop:bayes}
emphasizes a systematic discrepancy at order $1/t$ between ML prediction
and Bayesian prediction, which is often more ``centered'' as in
Laplace's rule.

This is characterized by a canonical vector
field on a statistical manifold indicating the direction of the
difference between
ML and Bayesian predictors, as follows. In intrinsic terms,
the posterior mean~\eqref{eq:posterior} in Proposition~\ref{prop:bayes}
is\footnote{The
equality between \eqref{eq:posterior} and \eqref{eq:intrinsicpost} holds
only at $\thetaML_t$; the value of \eqref{eq:posterior} is not intrinsic
away from $\thetaML$. The equality relies on $\partial_\theta L=0$ at
$\thetaML$ to cancel curvature contributions.}
\begin{equation}
\label{eq:intrinsicpost}
f(\thetaML)-
\frac1t
(\nabla^2 L)^{-1}\left(\d f,\d\ln
\frac{\pi}{\sqrt{\det(- \nabla^2 L)}}\right)
- \frac1{2t} \Tr \left((\nabla^2L)^{-1} \nabla^2 f\right)+O(1/t^2)
\end{equation}
where $L(\theta)=\sum_{i=t}^t\ln p_\theta(x_i)$ as above and where $\nabla^2$ is the Riemannian Hessian
with respect to any Riemannian metric on $\theta$, for instance the
Fisher metric. This follows
from a direct Riemannian-geometric computation (e.g., in normal
coordinates). 
In this expression both, the prior $\pi(\d\theta)$ and $\sqrt{\det(-
\nabla^2 L)}$ are volume forms on the
tangent space so that their ratio is coordinate-independent.\footnote{This is
clear when dividing both by the Riemannian volume form $\sqrt{\det g}$:
both the prior density $\pi/\sqrt{\det g}$ and $\sqrt{\det
(-g^{-1}\nabla^2 L)}$ are intrinsic.}

At first order in $1/t$, this is
the average of $f$ under a Riemannian
Gaussian distribution\footnote{i.e., the image by the exponential map of
a Gaussian distribution in a tangent plane.} with covariance matrix $\frac1t (-\nabla^2
L)^{-1}$, but centered at 
$\thetaML-\frac1t (\nabla^2 L)^{-1} \d\ln (\pi/\sqrt{\det(-\nabla^2 L)})$
instead of $\thetaML$.

Thus, if we want to approximate the posterior Bayesian distribution
by a Gaussian, there is a systematic shift $\frac1t V(\thetaML)$ between
the ML estimate and the center of the Bayesian posterior, where $V$ is
the data-dependent vector field
\begin{equation}
V\deq -(\nabla^2 L)^{-1} \d \ln \left(\pi/\sqrt{\det (-\nabla^2 L)}\right)
\end{equation}
A particular case is when $\pi$ is the Jeffreys prior: then 
\begin{equation}
\label{eq:VJef}
V=\frac12
(\nabla^2 L)^{-1} \d \ln \det (-\I^{-1}\nabla^2 L)
\end{equation}
is an
intrinsic vector field defined on any statistical manifold, depending on
$x_{1:t}$.

\begin{prop}
\label{prop:V}
When the prior is the Jeffreys prior,
the vector $V$ is 
\begin{equation}
V^i=\frac12 (\nabla_i\nabla_j L)^{-1}(\nabla_k\nabla_l L)^{-1}\,\nabla_j
\nabla_k \nabla_l L
\end{equation}
in Einstein notation,
where $L(\theta)=\frac1t\sum_{s=1}^t\ln p_\theta(x_s)$ is the
log-likelihood function, and $\nabla$ is the Levi-Civita connection of the
Fisher metric.\footnote{Note that $\nabla_j\nabla_k\nabla_l
L$ is not fully symmetric. Still it is symmetric at $\thetaML$, because the
various orderings differ by a curvature term applied to $\nabla L$ with
vanishes at $\thetaML$.}

If $p_\theta$ is an exponential family with the Jeffreys prior, the value of $V$ at
$\thetaML$ does not depend on the observations $x_{1:t}$ and is equal to
\begin{equation}
V^i(\thetaML)=\frac14\I^{ij}\I^{kl} T_{jkl}
\end{equation}
where $T$ is the skewness tensor \cite[Eq.~(2.28)]{Amari2000book}
\begin{equation}
T_{jkl}(\theta)\deq \E_{x\sim p_\theta} 
\frac{\partial \ln p_\theta(x)}{\partial \theta^j}
\frac{\partial \ln p_\theta(x)}{\partial \theta^k}
\frac{\partial \ln p_\theta(x)}{\partial \theta^l}
\end{equation}
\end{prop}

$V(\thetaML)$ is thus an intrinsic, data-independent vector field for exponential
families, which characterizes the discrepancy between maximum
likelihood and the ``center'' of the Jeffreys posterior distribution.
Note that $V$ can be computed from log-likelihood derivatives only.
This could be useful for regularization of the ML estimator in
statistical learning.


\paragraph{Proofs (sketch).}
\label{sec:proofs}

\begin{dem}[ of Proposition~\ref{prop:onlineML}] Minimization of a Taylor
expansion of log-likelihood around $\thetaML_t$. This is justified
formally by applying the implicit function theorem to 
$F\from (\eps,\theta) \mapsto \partial_{\theta}\left(\eps \ln
p_\theta(x_{t+1})+\frac1t \sum_{i=1}^t \ln p_\theta(x_t)\right)$ at point
$(0,\thetaML)$.
\end{dem}

\begin{dem}[ of Proposition~\ref{prop:NML}]
Abbreviate $\theta_y\deq\thetaMLp{y}_{t}$. From
Proposition~\ref{prop:onlineML} we
have \begin{equation}\theta_y=
\thetaML_t+\frac1t J_t^{-1}\partial_\theta \ln p_\theta(y)+O(1/t^2)\end{equation}
and expanding $\ln p_\theta(y)$ around $\thetaML_t$ yields
$
p_{\theta_y}(y)=
p_{\thetaML_t}(y)(1+\transp{(\theta_y-\thetaML_t)}\partial_\theta
\ln p_{\theta}(y))
+O((\theta_y-\thetaML)^2)
$
and
plugging in the value of $\theta_y-\thetaML_t$ yields the result.
\end{dem}

\begin{dem}[ of Proposition~\ref{prop:bayes}]
The posterior mean is $(\int
f(\theta)\alpha(\theta)p_\theta(x_{1:t})\d\theta)/(\int
\alpha(\theta)p_\theta(x_{1:t})\d\theta)$.
From \cite{TierneyKadane1986}, if $L_1(\theta)=\frac1t \ln
p_\theta(x_{1:t})+\frac1t g_1(\theta)$ and $L_2=\frac1t \ln
p_\theta(x_{1:t})+\frac1t g_2(\theta)$ we have
\begin{equation}
\label{eq:TK}
\frac{\int e^{tL_2(\theta)}\d\theta}{\int e^{tL_1(\theta)}\d\theta}=
\sqrt{\frac{\det H_1}{\det
H_2}}\,e^{t(L_2(\theta_2)-L_1(\theta_1))}(1+O(1/t^2))
\end{equation}
where $\theta_1=\argmax L_1$, $\theta_2=\argmax L_2$, and $H_1$ and $H_2$
are the Hessian matrices of $-L_1$ and $-L_2$ at $\theta_1$ and
$\theta_2$, respectively. Here we have $g_1=\ln \alpha(\theta)$ and
$g_2=g_1+\ln f(\theta)$ (assuming $f$ is positive; otherwise, add a
constant to $f$).

From a Taylor expansion of $L_1$ as in Proposition~\ref{prop:onlineML} we find $\theta_1=\thetaML_t+\frac1t
J_t^{-1}\partial_\theta g_1(\thetaML_t)+O(1/t^2)$ and likewise for
$\theta_2$. So $\theta_1-\theta_2=\frac1t
J_t^{-1}\partial_\theta (g_1-g_2)(\thetaML_t)+O(1/t^2)$.
Since $\theta_2$ maximizes $L_2$, a Taylor expansion of $L_2$
around $\theta_2$ gives
\begin{align}
L_2(\theta_1)&=L_2(\theta_2)-\frac12\transp{(\theta_1-\theta_2)}H_2(\theta_1-\theta_2)+O(1/t^3)
\end{align}
so that, using $L_2=L_1+\frac1t \ln f$ we find
\begin{align}
L_2(\theta_2)-L_1(\theta_1)&=
L_2(\theta_1)-L_1(\theta_1)+\frac12\transp{(\theta_1-\theta_2)}H_2(\theta_1-\theta_2)+O(1/t^3)
\\&=\frac1t \ln f(\theta_1)+\frac1{2t^2} \transp{(\partial_\theta\ln
f)}J_t^{-1}H_2J_t^{-1}\,\partial_\theta\ln f+O(1/t^3)
\end{align}
where the second term is evaluated at $\thetaML_t$. We
have $H_2=J_t+O(1/t)$, so
$\exp(t(L_2(\theta_2)-L_1(\theta_1)))=f(\theta_1)(1+\frac1{2t}
\transp{(\partial_\theta\ln
f)}J_t^{-1}\,\partial_\theta\ln f+O(1/t^2))$. Meanwhile, by a Taylor
expansion of $\ln \det (-\partial^2_\theta L_2(\theta_2))$ around $\theta_2$,
\begin{align}
\det H_2=
\det (-\partial^2_\theta L_2(\theta_2))
&=\det (
-\partial^2_\theta
L_2(\theta_1))\left(1+\transp{(\theta_2-\theta_1)}\partial_\theta\ln\det
(-\partial^2_\theta
L_2)+O(\theta_2-\theta_1)^2\right)
\end{align}
and from $L_2=L_1+\frac1t \ln f$ and $\det(A+\eps
B)=\det(A)(1+\eps\Tr(A^{-1}B)+O(\eps^2))$,
\begin{align}
\det (-\partial^2_\theta L_2(\theta_1))&=\det(-\partial^2_\theta
L_1(\theta_1))\left(1+\frac1t \Tr\left((\partial^2_\theta
L_1)^{-1}\partial^2_\theta(\ln
f)\right)+O(1/t^2)\right)
\\&=(\det H_1)\left(1-\frac1t \Tr\left(H_1^{-1}\partial^2_\theta(\ln
f)\right)+O(1/t^2)\right)
\end{align}
so, collecting,
\begin{align}
\sqrt{\frac{\det H_1}{\det H_2}}&=
1-\frac12 \transp{(\theta_2-\theta_1)}\partial_\theta \ln \det
(-\partial^2_\theta L_2)
+\frac1{2t} \Tr \left(H_1^{-1}\partial^2_\theta(\ln f)\right)+O(1/t^2)
\end{align}
but $\theta_2-\theta_1=J_t^{-1}\partial_\theta \ln f+O(1/t^2)$, and
$L_2=L+O(1/t)$ and $H_1=J_t+O(1/t)$, so that
\begin{align}
\sqrt{\frac{\det H_1}{\det H_2}}&=
1-\frac1{2t}
\transp{(\partial_\theta \ln f)}J_t^{-1}\partial_\theta
\ln\det (-\partial^2_\theta L)+ \frac1{2t} \Tr \left(J_t^{-1}\partial^2_\theta(\ln f)\right)
+O(1/t^2)
\end{align}

Collecting from \eqref{eq:TK}, 
expanding $f(\theta_1)
= f(\thetaML_t)(1+\frac1t
\transp{(\partial_\theta \ln f)}J_t^{-1}\partial_\theta\ln
\alpha+O(1/t^2))$, and expanding $\partial_\theta\ln f$ in terms of
$\partial_\theta f$ proves Proposition~\ref{prop:bayes}.
\end{dem}

\begin{dem}[ of Corollary~\ref{cor:exp}]
Let us work in natural coordinates for an
exponential family (indeed, since the statement is intrinsic, it is enough to
prove it in some coordinate system). In these coordinates, for any $x$,
$\partial^2_\theta \ln p_\theta(x)=-\I(\theta)$ with $\I$ the Fisher
matrix, so that $-\partial_\theta^2 L=\I(\theta)$. Apply
Proposition~\ref{prop:bayes} to $f(\theta)=p_\theta(y)$, expanding
$\partial_\theta f=f\partial_\theta \ln f$ and using $\partial_\theta^2
\ln f=-\I(\theta)$.
\end{dem}


\begin{dem}[ of Proposition~\ref{prop:V}]
The Levi-Civita connection on a Riemannian manifold with metric $g$
satisfies $\nabla_l \ln \det A_i^j=(A^{-1})^i_j\nabla_l A^j_i$
thanks to $\partial \ln \det M=\Tr(M^{-1}\partial M)$ and by expanding
$\nabla A$. Applying this to $A^j_i=\I^{jk}\nabla^2_{ki} L$ and using
$\nabla\I=0$ proves the
first statement.
Moreover, for any function $f$,
\emph{at a critical point of $f$},
$
\nabla_l\nabla_j\nabla_k f=\nabla_l \partial_j\partial_k f-\Gamma_{jk}^i
\nabla_l\nabla_i f
$
and consequently at a critical point of $f$, with
$H_{ij}=\nabla_i\nabla_j f$,
\begin{equation}
\nabla_l \ln \det (g^{ij}H_{jk})=
(H^{-1})^{ij}\nabla_l \partial_i\partial_j f
-(H^{-1})^{jk}\Gamma^i_{jk}H_{il}
\end{equation}

In the natural parametrization of an exponential family, $-\partial^2 L$
is identically equal to the Fisher metric $\I$.
Consequently, $\nabla_l \ln \det (-\I^{ij}\nabla_{jk}^2L)=\I^{ij}\nabla_l
\I_{ij}-\I^{jk}\Gamma^i_{jk}\I_{il}=-\I^{jk}\Gamma^i_{jk}\I_{il}$ since
$\nabla \I=0$.
So from~\eqref{eq:VJef}, using $d=\nabla=\partial$ for scalars, and $\nabla^2L=-\I$ at
$\thetaML$, we get in this parametrization
\begin{align}
V^m&=-\frac12 \I^{ml}\partial_l \ln \det (-\I^{-1}\nabla^2 L)
=\frac12  \I^{ml}\I^{jk}\Gamma^i_{jk}\I_{il}
=\frac12 \I^{jk}\Gamma^m_{jk}
\end{align}

The Christoffel symbols $\Gamma$ in this parametrization can be computed from
\begin{align}
\partial_i \I_{jk}(\theta)&=\partial_i \E_{x\sim p_{\theta}}
\partial_j \ln p_\theta(x)\partial_k \ln p_{\theta}(x)
\\&=T_{ijk}-\I_{ij}\E_{x\sim p_{\theta}}\partial_k \ln p_{\theta}(x)
-\I_{ik}\E_{x\sim p_{\theta}}\partial_j \ln p_{\theta}(x)
=T_{ijk}
\end{align}
because $\partial_i\partial_j \ln p_\theta(x)=-\I_{ij}(\theta)$ for any
$x$ in this parametrization, and because $\E \partial \ln p_\theta(x)=0$.
So
$\Gamma^i_{jk}=\frac12\I^{il}T_{jkl}$ in this parametrization.
This ends the proof.
\end{dem}

\paragraph*{Acknowledgments.} I would like to thank Peter
Grünwald for valuable comments.

\bibliographystyle{alpha}
\bibliography{nmlplugin}

\end{document}